\documentclass[aps, twocolumn, 8pt]{revtex4}
\usepackage{latexsym}
\usepackage{graphicx}

\usepackage{amssymb}
\def\II{\mathbb I}
\def\CC{\mathbb C}
\def\RR{\mathbb R}
\def\ZZ{\mathbb Z}

\begin{document}

\title{Implementable Quantum Bit-String Commitment Protocol}
\author{Toyohiro Tsurumaru}
\affiliation{Mitsubishi Electric Corporation,\\
Information Technology R\&D Center\\
5--1--1 Ofuna, Kamakura-shi, Kanagawa,
247-8501, Japan}

\begin{abstract}
Quantum bit {\it string} commitment[A.Kent, {\it Phys.Rev.Lett.}, {\bf 90}, 237901 (2003)] or QBSC
is a variant of bit commitment (BC).
In this paper, we propose a new QBSC protocol
that can be implemented using currently available technology,
and prove its security
under the same security criteria as discussed by Kent.
QBSC is a generalization of BC, but has slightly weaker requirements,
and our proposed protocol is {\it not} intended to break the no-go theorem of quantum BC.

\vskip10pt
PACS number(s): 03.67.-a, 03.67.Dd, 89.70.+c
\end{abstract}

\maketitle

\section{Introduction}
Quantum bit string commitment (QBSC) was proposed by Kent\cite{Kent},
as a generalization of bit commitment (BC).
The fundamental goal of QBSC is the same as that of BC;
that is, the sender first sends evidence of assigned data
over a communication channel without revealing the actual data.
Then, after an interval,
the sender reveals the data and the receiver verifies that the data have not been modified.

The difference in QBSC is that
the sender commits a string $A=(a_1,\cdots,a_n)$ with $n>1$ in a single protocol session,
and a limited (but certain) number of its bits are accessible by the receiver before the open phase.
This situation is referred to as $(m,n)$ {\it bit string commitment} in Ref.\cite{Kent},
where $m$ $(\le\! n)$ is the maximum number of accessible bits.
Conventional BC schemes correspond to the case where $n=1$ and $m=0$,
and in this sense QBSC is a generalization of BC.

In this paper, we propose a new QBSC protocol that can be implemented
using currently available technology, such as single-photon sources and detectors,
and then prove its security.
The main difference from the previous protocol is in the quantum measurement procedure
performed by the receiver to verify a commitment.
In our scheme, the receiver is allowed to perform the measurements in the commitment phase using randomized bases,
which means that it is not necessary to preserve the quantum state until the open phase.

Although QBSC is a generalization of BC,
the security level that it achieves is slightly weaker than that of BC unless $m=0$.
We would like to stress that our proposal is {\it not} intended
to break the no-go theorem of Lo-Chau\cite{Lo} and Mayers\cite{Mayers},
which states that if one takes full advantage of quantum computers
and quantum communication channels, any type of nonrelativistic BC scheme can be attacked\footnote{%
On the contrary, a bit commitment scheme that exploits special relativity
has been proposed and the security proof has been given.
See Ref.\cite{Kent2}.}.
Our ultimate goal is to devise useful and secure quantum cryptographic schemes that do not rely on BC.

The no-go theorem of BC often makes people think
that it might be impossible to create any useful cryptographic protocols based on quantum theory,
except quantum key distribution (QKD).
This is mainly because BC is one important building block in a whole list of interesting tools,
such as multi-party protocols or zero-knowledge proof, rooted in classical cryptology
(see {\it e.g.}, Ref.\cite{Delfs}\cite{Goldreich}).
However, as pointed out in Ref.\cite{Kent}, we are not yet able even to characterize
the range of cryptographic tasks for which perfectly secure quantum protocols might possibly exist.
Thus, we assume here that there must be distinct security notions that are unique to quantum cryptography,
and seek protocols that do not necessarily have a classical counterpart.

This paper is organized as follows.
In Section \ref{sec:previous_method},
we briefly review some of the results by Kent\cite{Kent},
including the basic structure of QBSC and its security criteria.
Subsequently, we describe the proposed protocol in Section \ref{sec:proposed_protocol}
and prove its security in Section \ref{sec:security_evaluation}.
Then in Section \ref{sec:implementation},
we briefly comment on the implementation and finally conclude in Section \ref{sec:conclusion}.

\section{Previous Method}\label{sec:previous_method}
The original QBSC was proposed in Ref.\cite{Kent}, with two example schemes, called Protocol 1 and Protocol 2.
Before presenting our protocol,
we briefly review the results by Kent in this section.
Both of the example protocols share a simple basic structure, as follows.
Throughout this paper,
we will call the sender Alice and the receiver Bob.

\subsection{Basic Structure of Kent's Scheme}
\subsubsection{Procedure}
Alice and Bob proceed as follows.

\ 

{\bf Commitment Phase}
\begin{enumerate}
\item Alice chooses a bit string $A=(a_1,\dots,a_n)$ to be committed to
and sends Bob the corresponding state $|\Psi_A\rangle$.
\item Bob preserves the received state $|\Psi_A\rangle$.
\end{enumerate}

{\bf Open Phase}
\begin{enumerate}
\item Alice unveils $A$ to Bob.
\item Bob verifies $A$ by a projective measurement on $|\Psi_A\rangle$.
If the outcome is consistent with $A$, he accepts the commitment,
otherwise he rejects it.
\end{enumerate}
The exact form of the state $|\Psi_A\rangle$ differs depending on the type of protocol
and will be given below.
\subsubsection{Security Requirement}\label{sec:BC_requirement}
For BC, there are two security requirements, which are called the concealing and binding conditions\cite{Goldreich}.
Similar notions are also used for QBSC, although with somewhat relaxed restrictions\cite{Kent}.
\begin{itemize}
\item[-] {\bf Concealing Condition.}
The receiver can access only a limited number of bits
in the committed string $A$ before the open phase.

The number of accessible bits is referred to as $m$ in Ref.\cite{Kent}.
\item[-] {\bf Binding Condition.}
It is unlikely that Alice will change the content of her commitment $A$ after the commitment phase.
\end{itemize}

Note that for BC schemes,
Bob is assumed to gain no information whatsoever regarding committed bit $b$ until Alice unveils it,
whereas in QBSC
perfect concealment is dispensed with; 
Bob is allowed to extract a certain but limited amount of information from $A$.

The meaning of the term `unlikely' in the definition of the binding condition
differs in the two protocols.  This point will be discussed below.

As to the possibility of each player cheating,
Alice cheating puts the binding at risk and Bob cheating corresponds to concealment.
Throughout this paper,
we consider only cases where either one of them is cheating, but not both.
Thus, the binding and concealing conditions can be discussed separately.

\subsubsection{Measurement by Bob}
When considering Alice's cheating,
we denote the quantum state sent to Bob as a density matrix $\rho$,
since in general, Alice would be likely to use a randomized strategy or send an entangled state.
Note here that whatever strategy Alice follows,
once she transmits the quantum state to Bob, the density matrix $\rho$ is fixed.
Honest Bob verifies Alice's commitment
by performing a projective measurement of $\rho$
using an orthonormal basis that includes $|\Psi_A\rangle$.
Bob then accepts it as a correct commitment of string $A$ with
the probability
\begin{equation}
{\rm Pr}[A|\rho]=\langle\Psi_A|\rho|\Psi_A\rangle\ .
\label{eq:def_prob}
\end{equation}
\subsection{Example Protocols}
\subsubsection{Protocol 1}
Define qubit states
\[\psi_0=|0\rangle,\ \psi_1=\sin\theta|0\rangle+\cos\theta|1\rangle\ .\]
Alice sends particles in the states $\psi_{a_1},\dots,\psi_{a_n}$
for $A=(a_1,\dots,a_n)$, $a_i\in\{0,1\}$.
In other words, she transmits to Bob the state $|\Psi_A\rangle := |\psi_{a_1}\rangle\otimes\cdots\otimes|\psi_{a_n}\rangle$.
Here $\theta\in\RR$ is a constant that is determined according to the security parameter.

\paragraph{Concealing Condition}
From the Holevo bound\cite{Nielsen},
the information $m$ that is accessible by Bob is bounded by
\begin{equation}
m\le S(\rho)=\left[H_2\left(\frac{1+\sin\theta}2\right)\right]^n\ ,
\label{eq:Kent_protocol1_1}
\end{equation}
where $H_2(x)=-x\log x -(1-x)\log(1-x)$.
Hence $m$ can be arbitrarily small by adjusting $\theta$.
\paragraph{Binding Condition.}
In Protocol 1,
binding is discussed for each bit separately.
Let $p^j_i=\langle\psi_j|\rho_i|\psi_j\rangle$ be the probability of Bob accepting
a revelation of $j$ for the $i$th bit.
We have
\begin{equation}
p_i^0+p_i^1\le \cos^2[(\pi-2\theta)/4]+\sin^2[(\pi+2\theta)/4]\ ,
\label{eq:Kent_protocol1_2}
\end{equation}
where the RHS can again be chosen arbitrarily close to 1 by choosing suitable $\theta$.
In general, the smaller $\theta$ is, the stronger the binding becomes;
simultaneously, however, concealing becomes weaker, as can be seen from Inequality (\ref{eq:Kent_protocol1_1}).
Thus, $\theta$ needs to be chosen with this balance in mind.

\subsubsection{Protocol 2}
In Protocol 1,
as $m$ in (\ref{eq:Kent_protocol1_2}) is in the same order as $n$,
most bits from $A$ are accessible by Bob,
because the amount of data encoded per qubit is rather small.
Thus, another protocol at the other extreme with a much 
higher encoding rate is considered here 
.

The basic idea here is that instead of using qubits,
one encodes $A$ into a general $D$-dimensional vector space ${\cal H}_D = \CC^D/\CC$.
Thus, one chooses $O\left(\exp({\rm const.}D)\right)$ states $\{|\Psi_A\rangle\}$
out of ${\cal H}_D$.

\paragraph{Binding Condition.}
In Protocol 1, we discussed bitwise security;
here we take a different approach.
\newtheorem{theorem}{Definition}
\begin{theorem}
Take an arbitrary value of $r\in \ZZ$ and $\varepsilon\in \RR$ $(\varepsilon,r>0)$.
A QBSC protocol is binding if, by choosing a large enough $n$,
the following inequality holds for arbitrarily chosen $A_1,\dots,A_r$
and $\rho$,
\begin{equation}
\sum_i {\rm Pr}[A_i|\rho]\le 1+\varepsilon\ .
\label{eq:def_security1}
\end{equation}
Here, $n$ is the length of the bit string that is to be committed.\hfill$\Box$
\end{theorem}
Intuitively, this definition can be illustrated as follows.
A cheating Alice might want to postpone her decision by selecting $x$
patterns of bit strings $A_1,\dots, A_x$ in the commitment phase.
Then, later in the open phase, she could unveil any desired string
among them.
In such a case, the maximum probability of her success is less than $1/x$ if $x<r$,
and is less than $1/r$ if $x>r$.

In fact, it can be shown\cite{Kent} that
if the committed states $|\Psi_A\rangle$ are chosen to be approximately orthogonal:
\begin{equation}
|\langle\Psi_A|\Psi_{A'}\rangle|\le\sin\theta=\epsilon\ \ll1\ \ {\rm for}\ \ \forall A\ne \forall A'
\label{eq:Kent_protocol2_1}
\end{equation}
the above binding condition (\ref{eq:def_security1}) is true for
$\varepsilon\le (r-1)\epsilon$.
That is,
if one can construct approximately orthogonal states $|\Psi_A\rangle$,
the binding condition is automatically guaranteed.
\paragraph{Approximately Orthogonal States.}
In this protocol, one needs to choose $\exp({\rm const.}D)$ states $|\Psi_A\rangle$
out of $D$-dimensional vector space while keeping approximate orthogonality
as in (\ref{eq:Kent_protocol2_1}).
Although this seems impossible at first sight,
interestingly, it can in fact be easily achieved by using classical error-correcting codes\cite{Buhrman}.
One example is as follows.

Take an error-correcting code $E:\{0,1\}^n\to \{0,1\}^m$ with
information rate $R=n/m$ and minimum distance $d$ and let $\delta=d/m$.
Next choose quantum states $|h_A\rangle\in\CC^{2m}/\CC$ corresponding to string $A\in\{0,1\}^n$
as:
\begin{equation}
|h_A\rangle:=\frac1{\sqrt{m}}\sum_{i=1}^m|i\rangle\otimes|E_i(A)\rangle\ .
\label{eq:Kent_protocol2_2}
\end{equation}
Then, as $|\langle h_A|h_{A'}\rangle|\le (1-\delta)$ for $A\ne A'$,
an arbitrarily small $\epsilon$ can be chosen by defining $|\Psi_A\rangle$ as
$|\Psi_A\rangle:=|h_A\rangle\otimes\cdots\otimes|h_A\rangle$.

\section{Proposed Protocol}\label{sec:proposed_protocol}

The protocols described in the previous section are certainly significant,
in that they achieve a completely new type of security by using quantum mechanics.
However, it seems almost impossible to implement them using current technology.
There are two main reasons.
\begin{enumerate}
\item
Bob needs to preserve the quantum state $|\Psi_A\rangle$ as it was created by Alice,
until the open phase.
\item
In Protocol 2,
highly entangled states, such as $|h_A\rangle$ given in (\ref{eq:Kent_protocol2_2}),
are necessary.
\end{enumerate}
In particular, the first point seems the more difficult to overcome.
Thus, we propose a modified version of QBSC that can be implemented,
and subsequently prove its security.

In Kent's protocol,
Bob needs to preserve the quantum state $\rho$ as
it was sent from Alice
in the commitment phase until the open phase. 
This is because Bob verifies $\rho$ by projective measurements
using an orthonormal basis including $|\Psi_A\rangle\langle\Psi_A|$, after Alice has unveiled string $A$.
Therefore,
we cosider here whether Bob can distinguish $\rho$ sufficiently accurately without knowing $A$,
because if that is the case, Bob does not need to preserve $\rho$ for a long period.

In order to actually do this,
we change Bob's verification procedure.
That is, Bob measures $\rho$ as soon as he receives it
using randomly chosen orthogonal bases,
and then stores the result until the open phase.
Although much less information is obtained in this way 
than is obtained by directly comparing $\rho$ with $|\Psi_A\rangle\langle\Psi_A|$,
it is enough to detect cheating by Alice, as we will show in Section \ref{sec:security_evaluation}.

\begin{figure*}[t]
\begin{center}
\includegraphics[trim=2cm 10.5cm 2cm -7.0cm, clip, scale=0.5]{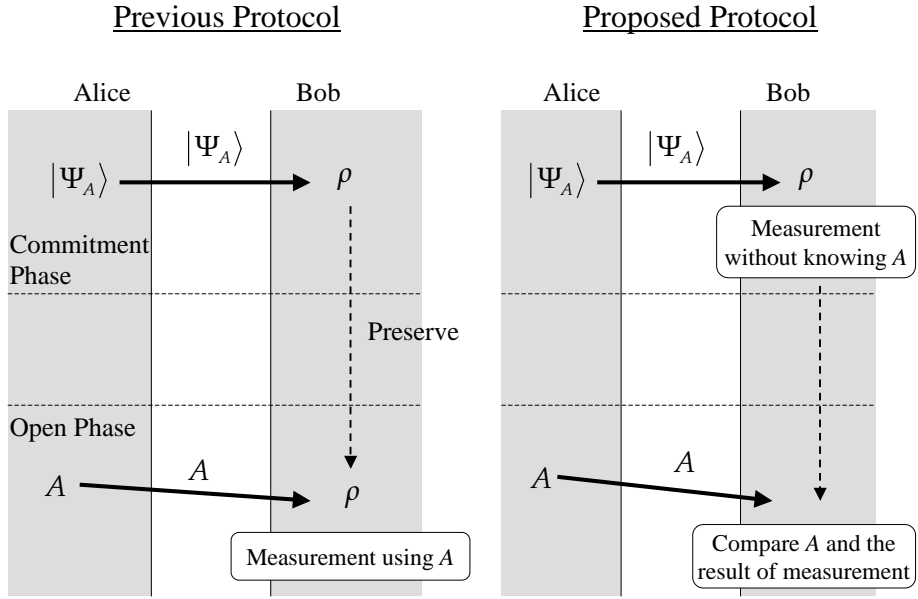}
\end{center}
\caption{Differences between the previous and proposed protocols.}
\label{fig:proposal}
\end{figure*}

\subsection{Quantum State for Encoding}
First we define the quantum state $|\Psi_A\rangle$,
which is used as a commitment of the string $A$.

\subsubsection{Error-Correcting Code}
Throughout this section,
we assume that string $A$ is not necessarily a bit string but is in general a
$q$-ary string $A=(a_1,\dots,a_{n'}),\ a_i\in\{0,\dots,q-1\}$.
If the original data are a bit string,
one needs to transform them into a $q$-ary string using some appropriate surjective mapping.
In this case, $A$ stands for the data of $n=\lceil n'\log_2 q\rceil$ bits.

Then we fix a $q$-ary classical $(N,n',d)$ error-correcting code :
\[E:A\mapsto E(A)=(e_1,\dots,e_N)\ ,\ \ e_i\in\{0,\dots,q-1\}\ .\]
The sender stretches the string $A$ into a $q$-ary string $E$ of length $N$
using this error-correcting code.

\subsubsection{Quantum States of `Particles'}
The commitment $|\Psi_A\rangle$ takes the form
\begin{equation}
|\Psi_A\rangle=|\Psi_{E(A)}\rangle:=|e_1\rangle\otimes\cdots\otimes|e_N\rangle\ ,
\label{eq:define_Psi_A_by_particle}
\end{equation}
where each $|e_i\rangle$ is a $D$-dimensional vector.
That is,
the sender commits by sending $N$ `particles' having
internal degrees of freedom in a $D$-dimensional vector space ${\cal H}_D=\CC^D/\CC$.
Each $q$-ary value $e_i$ is encoded into this $D$-dimensional space of
the $i^{\rm th}$ particle.

We use the same set of $D$-dimensional states for each particle and denote them as
\begin{equation}
|0\rangle,\cdots,|q-1\rangle\ \in{\cal H}_D\ .
\label{eq:def_particle_state}
\end{equation}
Hence, Eqn.(\ref{eq:define_Psi_A_by_particle}) means that
the $i^{\rm th}$ particle has $e_i^{\ {\rm th}}$ states
of (\ref{eq:def_particle_state}).
The $q$ states of (\ref{eq:def_particle_state}) are all distinct
but do not necessarily form an orthonormal basis.

It is assumed here that $D$ satisfies $q=lD$ for $l\in\ZZ$,
and that the states of (\ref{eq:def_particle_state})
can be grouped together into $l$ types of orthonormal basis $M(i)$,
$i=0,\dots,l-1$
\begin{equation}
M(i):=\{|i;0\rangle,\dots,|i;D-1\rangle\},\ \ \langle i;j|i;k\rangle = \delta_{jk}
\label{eq:def_smaller_vector}
\end{equation}
with
\[|iD+j\rangle=|i;j\rangle\ .\]
Also, for the sake of simplicity,
we assume that the orthogonal bases $M(i)$ are symmetric under a finite group $G$
that acts on ${\cal H}_D$.
In other words, for any $0\le i\le l-1$, $g\in G$,
there exists $0\le i'\le l-1$ satisfying
\begin{equation}
gM(i) := 
\{R(g)|i;0\rangle,\ \dots,\ R(g)|i;D-1\rangle\}=M(i')\ ,
\label{def:symmetric}
\end{equation}
where $R$ is a representation of $G$ in ${\cal H_D}$.
In Eqn.(\ref{def:symmetric}), states are identified if they differ only up to a phase
of complex number.

\subsection{Protocol}\label{sec:proposed_protocol_procedure}
Alice and Bob proceed as follows.

\ 

{\bf Commitment Phase}
\begin{enumerate}
\item Alice chooses a $q$-ary string $A=(a_1,\dots,a_{n'})$ to which she is committed
and computes the corresponding codeword $E(A)=(e_1,\dots,e_N)$.
\item Bob chooses a $l$-ary random string

$S=(s_1,\dots,s_N)\in\{0,\dots,l-1\}^N$.
\item Alice sends $N$ particles $|e_1\rangle,\dots,|e_N\rangle$ to Bob.
\item Bob performs measurements on each received particle $|e_i\rangle$
with the basis $M(s_i)$ and records the result.
\end{enumerate}

{\bf Open Phase}
\begin{enumerate}
\item Alice reveals $A$ to Bob.
\item Bob computes the codeword $E(A)$.
\item Bob verifies the following condition for each particle $1\le i\le N$\ :
\begin{itemize}
\item[-]
If the basis that he chose is correct, {\it i.e.}, $|e_i\rangle\in M(s_i)$,
he obtains $|e_i\rangle$ as a result of his measurement.
\end{itemize}
\item
If the above condition does not hold for any $i$,
Bob rejects the commitment, otherwise, he accepts it.
\end{enumerate}

\subsection{Differences from Kent's Protocols}
As stated earlier,
Bob does not need to preserve the quantum state $\rho$ until the open phase
as he examines $\rho$ in the commitment phase.
By using randomized bases, he selects correct bases in a probabilistic way,
and is therefore able to verify Alice's commitment confidently.

In addition, players only need to be able to create and detect $D$-dimensional
quantum states, and thus the highly entangled states, {\it e.g.}, $|h_A\rangle$ given in (\ref{eq:Kent_protocol2_2}),
are unnecessary.
For instance, the case of $D=2$ can be implemented using single-photons.

Error-correcting code is used here to
ensure approximate orthogonality of states $|\Psi_{E(A)}\rangle$ as in Kent's protocol
({\it c.f.}, Eqn.(\ref{eq:Kent_protocol2_1})).
Indeed in our protocol,
Alice sends $|\Psi_{E(A)}\rangle:=|e_1\rangle\otimes\cdots\otimes|e_N\rangle$,
which satisfies
\begin{equation}
\forall A\forall A'\left(
A\ne A'\to\left|\left\langle \Psi_{E(A)}| \Psi_{E(A')}\right\rangle\right|
<\bar\beta^d\right)\ .
\label{eq:orthogonality_proposed_protocol}
\end{equation}
Here $\bar\beta$ is the maximum value of the inner product of the particle states
defined in (\ref{eq:def_particle_state}),
{\it i.e.}, $|\langle e|e'\rangle|\le\bar\beta$ for $\forall e\ne e'$
(see also Paragraph \ref{par:orthogonality_of_VEA_i}).
In the next section, we will exploit this property to prove the binding condition.

\section{Security Evaluation}\label{sec:security_evaluation}
Now we will prove the security of the proposed scheme.
Throughout this section,
we consider an ideal case with a noiseless channel and error-free detection.
(For non-ideal cases, see Section \ref{sec:implementation}.)

As to what each player can do in quantum protocols,
we follow the definition given by Yao\cite{Yao} and Lo-Chau\cite{Lo2}.
That is, quantum protocol is formalized in a Hilbert space
${\cal H}_{\rm T}={\cal H}_{\rm A}\otimes{\cal H}_{\rm B}\otimes{\cal H}_{\rm C}$,
where ${\cal H}_{\rm A}$ (resp. ${\cal H}_{\rm B}$) refers to the space in which Alice (resp. Bob) can operate.
${\cal H}_{\rm C}$ is the communication channel.
Every step of the protocol is done by unitary operations by Alice on ${\cal H}_{\rm A}\otimes{\cal H}_{\rm C}$
and those by Bob on ${\cal H}_{\rm B}\otimes{\cal H}_{\rm C}$, alternately.

\subsection{Concealing condition}
In the proposed protocol, $n\log_2 q$-bit data are encoded in $N\log_2 D$ qubits.
Thus, the number of bits accessible by Bob, $m$, satisfies $m\le N\log_2 D$ owing to the Holevo bound.
Hence, the protocol is concealing when the bit number of $A$ is greater than the qubit number,
or $N\log D<n\log q$.

\subsection{Binding Condition}\label{sec:proposal_binding}
In the remainder of this section, we will prove the following theorem.
\newtheorem{theorem1}[theorem]{Theorem}
\begin{theorem1}
The proposed protocol is binding in terms of Definition 1.
\end{theorem1}

\subsubsection{Basic Idea of the Proof}
The basic idea of the proof is the same as given in Ref.\cite{Kent}.
Namely, we exploit the approximate orthogonality (\ref{eq:orthogonality_proposed_protocol}) of
$|\Psi_A\rangle$ (see Eqn.(\ref{eq:Kent_protocol2_1}) and the discussion nearby).
However, there is a complication, owing to the difference in the form of the probability ${\rm Pr}[A_i|\rho]$.
In our protocol, ${\rm Pr}[A_i|\rho]$ takes a different form,
as the receiver performs measurements on a randomly chosen orthogonal basis,
whereas in Kent's protocol, the receiver can make use of the string $A$ unveiled by Alice.

\paragraph{Form of ${\rm Pr}[A|\rho]$.}
The probability ${\rm Pr}[A|\rho]$ is given as follows.
First, for the sake of simplicity, suppose that $N=1$, that is,
suppose that Alice sends only one particle.
Then Bob chooses the correct basis for measurement
(satisfying $e_1(A)\in M(s_1)$) with probability $1/l$,
in which case he can confidently reject Alice's commitment
if the outcome is different from $|e_1(A)\rangle$.
On the other hand, if he chooses a wrong basis,
he may accept Alice's commitment no matter what he obtains as a result of his measurement.
Thus, averaged over the random variable $S$,
the acceptance rate by Bob takes the form
\begin{equation}
{\rm Pr}_{N=1}[A|\rho]:=\left(1-\frac1l\right)+\frac1l\langle e_1|\rho |e_1\rangle
={\rm Tr}\rho \pi(e_1)\ ,
\end{equation}
where
\begin{eqnarray}
\pi(e)&:=&\left(1-\frac1l\right)\II_D+\frac1l|e\rangle\langle e|\nonumber\\
&=&\left(1-\frac1l\right)\II_D+\frac1l|i;j\rangle\langle i;j|=:\pi(i;j)\ .
\label{eq:def_pi_e}
\end{eqnarray}
Here $\II_D$ is a unit matrix in ${\cal H}_D$.

Similarly for $N>1$, {\it i.e.},
if Alice sends more than one particle,
the acceptance rate takes the form
\[{\rm Pr}[A|\rho] = {\rm Tr}\rho P_{E(A)}\]
with
\begin{equation}
P_E=\pi\left(e_1\right)\otimes\cdots\otimes\pi\left(e_N\right)\ .
\label{eqn:def_Pi_E}
\end{equation}
\paragraph{What We Need to Prove.}
The sum of the acceptance rates for $A_1,\dots,A_r$,
which appears on the LHS of (\ref{eq:def_security1}),
takes the form
\begin{equation}
\sum_i{\rm Pr}[A_i|\rho]={\rm Tr}\left(\rho Q\right)
\label{eq:def_sum_prob}
\end{equation}
with
\begin{equation}
Q:=P_{E(A_1)}+\cdots+P_{E(A_r)}\ .
\label{eq:def_Q}
\end{equation}
With this operator $Q$,
what we need to do is to bound ${\rm Tr}\rho Q$ from above for an arbitrary density matrix $\rho$.
However, it is obviously sufficient
to consider instead the upper bound on $\langle\Psi|Q|\Psi\rangle$ for an arbitrary pure state $|\Psi\rangle$.
Moreover, in order to avoid complications from state normalizations,
we do not require $|\Psi\rangle$ to be normalized and instead discuss the upper bound of the quantity
$\langle\Psi|Q|\Psi\rangle/\langle\Psi|\Psi\rangle$.

\subsubsection{Vector Subspace $V(E(A_i))$}
In order to prove Theorem 2,
it is convenient to divide the $D^N$-dimensional space ${\cal H}_{D^N}$,
which consists of the degrees of freedom of the $N$ particle used for commitment.
We here focus especially on the eigenvectors of $P_{E(A_i)}$ with relatively
large eigenvalues,
as these vectors contribute most to
$\langle \Psi|P_{E(A_i)}|\Psi\rangle$ included in $\langle \Psi|Q|\Psi\rangle$.

\paragraph{Eigenstates of operator $P_E$.}
The eigenstate of $P_E$ with eigenvalue 1, which is the largest,
takes the form
\begin{eqnarray}
|\Psi_E\rangle&:=&|e_1\rangle\otimes\cdots\otimes|e_N\rangle\nonumber\\
&=&|i_1;j_1\rangle\otimes\cdots\otimes|i_N;j_N\rangle\ .
\end{eqnarray}
This becomes the state that Alice sends as her commitment of the $q$-ary string $A$,
if we let $E=(e_1,\dots,e_N)$ equal the codeword $E(A)$ of $A$.

It is evident that
all eigenvectors of $P_E$, including $|\Psi_E\rangle$
take the form
\begin{equation}
|\Psi_E;\Delta J\rangle
:=|i_1;j_1+\Delta j_1\rangle\otimes\cdots\otimes|i_N;j_N+\Delta j_N\rangle\ ,
\label{eq:def_Psi_E}
\end{equation}
since $P_E$ is a tensor product of the smaller operator $\pi(e_i)$'s defined in Eqn.(\ref{eq:def_pi_e}),
which act on particles.
Here $\Delta J=(\Delta j_1,\dots,\Delta j_N)$ is an arbitrary $D$-ary string,
and the sum $j+\Delta j$ appearing on the RHS of (\ref{eq:def_Psi_E}) is assumed to be in modulo $D$.
Recall that,
as defined in (\ref{eq:def_smaller_vector}),
for each $i$,
the vectors $|i;0\rangle,\dots,|i;D-1\rangle$ form an orthonormal basis in
the internal $D$-dimensional vector space ${\cal H}_D$ of a particle.
Hence, the states $|\Psi_E;\Delta J\rangle$ obviously form a complete orthonormal basis
of the vector space ${\cal H}_{D^N}$ formed by $N$ particles.

Roughly speaking,
$\Delta J$ indicates the difference of $|\Psi_E;\Delta J\rangle$ from $|\Psi_E\rangle$.
The eigenvalue of $|\Psi_E;\Delta J\rangle$ decreases exponentially
with the Hamming weight $HW(\Delta J)$, or the number of nonzero elements of the $\Delta J$:
\begin{equation}
P_E|\Psi_E;\Delta J\rangle = (1-1/l)^{HW(\Delta J)}|\Psi_E;\Delta J\rangle\ .
\end{equation}
For example, $|\Psi_E\rangle=|\Psi_E;\Delta J\rangle$ for $\Delta J=0$.

\paragraph{Vector subspace $V(E(A_i))$.}
Then we define $V(E(A_i))$ as a vector subspace of ${\cal H}_{D^N}$,
generated by eigenvectors $|\Psi_{E(A_i)};\Delta J\rangle$ with
$HW(\Delta J)<\alpha$.
The constant $\alpha$ here
is an integer that determines the size of the subspace
and is supposed to satisfy $0<\alpha<d$ (see Fig.\ref{fig:graphical}).
That is to say, any vector $|\Gamma_i\rangle\in V(E(A_i))$ can be expanded as
\begin{equation}
|\Gamma_i\rangle = \sum_{\{\Delta J\ |\ HW(\Delta J)<\alpha\}} w_{i,\Delta J}|\Psi_{E(A_i)};\Delta J\rangle
\label{eq:expansion2}
\end{equation}
The appropriate value of $\alpha$ will be discussed later.

\paragraph{$V(E(A_i))$'s are approximately orthogonal.}\label{par:orthogonality_of_VEA_i}
In fact the subspaces
$V(E(A_i))$'s with different values of $i$
are approximately orthogonal to each other.
Indeed, if we pick two elements of
vector basis $|\Psi_{E(A_i)};\Delta J\rangle$ and
$|\Psi_{E(A_j)};\Delta J'\rangle$
from different subspaces, {\it i.e.}, for $i\ne j$,
we have
\begin{equation}
\left|\langle\Psi_{E(A_i)};\Delta J|\Psi_{E(A_j)};\Delta J'\rangle\right|
<\bar{\beta}^{d-\alpha}\ ,
\label{eq:def_upper_bound_inner}
\end{equation}
where $\bar{\beta}$ is the upper bound on the inner products of
particle state $|0\rangle,\dots,|q-1\rangle$ defined in (\ref{eq:def_particle_state}).
In other words, $\bar{\beta} := \max_{e\ne e'}
\left|\langle e|e'\rangle\right|\ <1$.

Inequality (\ref{eq:def_upper_bound_inner}) follows because
when the two states appearing on the LHS are
represented using particle states as
$|\Psi_{E(A_i)};\Delta J\rangle=|e_1\rangle\otimes\cdots\otimes|e_n\rangle$
and
$|\Psi_{E(A_j)};\Delta J'\rangle=|e'_1\rangle\otimes\cdots\otimes|e'_n\rangle$,
$|e_i\rangle$ and $|e'_i\rangle$ are different for at
least $d-\alpha$ values of $i$,
owing to the minimum distance property of classical error correcting code.
This is the main reason for introducing classical error-correcting code
(see Fig.\ref{fig:graphical}).

\subsubsection{Decomposition of $|\Psi\rangle$}
\paragraph{Subspace $V$ and $V^\bot$.}
Next we define another subspace
\[V=V(E(A_1))+\cdots+ V(E(A_r))\ .\]
That is, any element of $V$ is given by the sum of the elements of
$V(E(A_1)),\dots,V(E(A_r))$.
As can be seen from the construction of $V(E(A_i))$,
$V$ is the subspace of the quantum message space ${\cal H}_{D^N}$
which contributes by far the most to $\langle\Psi|Q|\Psi\rangle$.

Then we decompose $|\Psi\rangle$ into $V$ and $V^\bot$,
where $V^\bot$ denotes the orthogonal complement of $V$.
\begin{eqnarray}
|\Psi\rangle :=|\Psi_V\rangle+|\Psi_\bot\rangle
&=& \sum_iv_i|\Gamma_i\rangle+|\Psi_\bot\rangle\ ,\label{eq:expansion1}\\
\sum_i|v_i|^2&=&1\label{eq:v_i_normalization}
\end{eqnarray}
with $|\Psi\rangle\in V$, $|\Psi_\bot\rangle\in V^\bot$ and $v_i\in\CC$.
The state 
$|\Gamma_i\rangle$ is assumed to belong to $V(E(A_i))$
and can thus be expanded as in Eqn.(\ref{eq:expansion2}).
We also assume that each $|\Gamma_i\rangle$ is normalized,
{\it i.e.},
the coefficients $w_{i,\Delta J}$ defined in Eqn.(\ref{eq:expansion2})
satisfy
\[\sum_{\{\Delta J\ |\ HW(\Delta J)<\alpha\}}|w_{i,\Delta J}|^2=1\ .\]
On the other hand, we do not require $|\Psi_\bot\rangle$ to be normalized.
Furthermore, 
$|\Psi_V\rangle$ is not a unit vector in general,
although the coefficient $v_i$'s is normalized as in Eqn.(\ref{eq:v_i_normalization}).
This is because
$|\Gamma_i\rangle$'s with different values of $i$ are not exactly orthogonal
as discussed above.
It should also be noted that
there is arbitrariness in the choice of $|\Gamma_i\rangle$
owing to this non-orthonognality.

\begin{figure}[t]
\begin{center}
\includegraphics[trim=2.7cm 12.5cm 2cm -7.0cm, clip, scale=0.37]{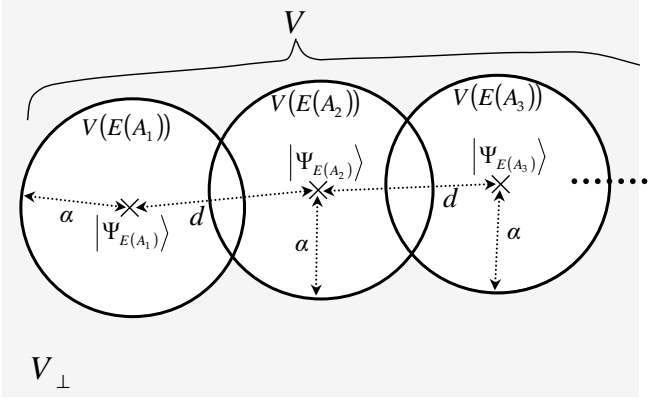}
\end{center}
\caption{A graphical image of vector spaces $V$ and $V^\bot$.
$V(E(A_i))$ is the vector subspace generated by $P_{E(A_i)}$'s
eigenstates with eigenvalues larger than $(1-1/l)^\alpha$.
Thus, the parameter $\alpha$ serves as the radius of each $V(E(A_i))$.
Parameter $d$ denotes the separation between each $V(E(A_i))$.
$V$ is the sum of all $V(E(A_i))$'s and $V^\bot$
is its orthogonal complement.}
\label{fig:graphical}
\end{figure}

It is still obvious that any vector
$|\Psi\rangle\in{\cal H}_{D^N}$ can be expanded as in Eqn.(\ref{eq:expansion1})
with appropriate rescaling.
\paragraph{$V^\bot$ contributes only little.}\label{par:contributes_little}
By definition,
$|\Psi_\bot\rangle$ satisfies
$\langle\Psi_V|\Psi_\bot\rangle=0$ for any $|\Psi_V\rangle\in V$,
and it is apparent that for any $i$,
$|\Psi_\bot\rangle\in \left[V(E(A_i))\right]^\bot$
(see Fig.\ref{fig:graphical}).
This means that $|\Psi_\bot\rangle$ can be expanded by
the eigenstates $|\Psi_{E(A_i)};\Delta J\rangle$ with Hamming weight
$HW(\Delta J)\ge \alpha$.

Thus, it immediately follows that
$\left\| P_{E(A_i)}|\Psi_\bot\rangle\right\|\le(1-1/l)^\alpha$
and we see that $|\Psi_\bot\rangle$ contributes little to
$\langle\Psi|Q|\Psi\rangle$ if $\alpha$ is sufficiently large.
See Lemma 3 for a more rigorous argument.

\subsubsection{Outline of the proof, and lemmas}
We have seen above that for large enough $d$ and $\alpha$
the subspace $V(E(A_i))$s are approximately orthogonal to each other,
and $|\Psi_\bot\rangle\in V^\bot$ contributes little
to $\langle\Psi|Q|\Psi\rangle$.
Then, the following inequalities should hold:
\begin{eqnarray}
\langle\Psi|Q|\Psi\rangle&=&\sum_i\langle\Psi|P_{E(A_i)}|\Psi\rangle\nonumber\\
&\simeq&\sum_i|v_i|^2\langle\Gamma_i|P_{E(A_i)}|\Gamma_i\rangle\le \sum_i|v_i|^2=1\ ,\label{eq:rough_ineq1}\\
\langle\Psi|\Psi\rangle&=&\langle\Psi_V|\Psi_V\rangle+\langle\Psi_\bot|\Psi_\bot\rangle
\simeq\sum_i|v_i|^2\langle\Gamma_i|\Gamma_i\rangle=1\ ,\nonumber\\
\label{eq:rough_ineq2}
\end{eqnarray}
which means $\langle\Psi|Q|\Psi\rangle/\langle\Psi|\Psi\rangle\simeq 1$.

In the subsequent subsection, we will use this idea to prove Theorem 2 rigorously,
but first we prepare two lemmas that are convenient.
First we show that the contribution of $|\Psi_\bot\rangle$ is indeed small.
\newtheorem{Lemma1}[theorem]{Lemma}
\begin{Lemma1}
\ 

For any $i$ and $s$, ($s\ge 0$, $s\in \ZZ$),
\begin{eqnarray}
\langle\Psi_\bot|\{P_{E(A_i)}\}^s|\Psi_\bot\rangle
&\le& (1-1/l)^{s\alpha}\langle\Psi_\bot|\Psi_\bot\rangle\ ,\label{lemma1_1}\\
\left\|P_{E(A_i)}|\Psi_\bot\rangle\right\|
&\le&(1-1/l)^\alpha \left\|\ |\Psi_\bot\rangle\right\|\ .\label{lemma1_2}
\end{eqnarray}
\end{Lemma1}
\noindent{\it Proof.}
(See also Paragraph \ref{par:contributes_little} and Fig.\ref{fig:graphical}.)
As $|\Psi_\bot\rangle\in V^\bot\subset \left[V(E(A_i)\right]^\bot$,
$|\Psi_\bot\rangle$ can be expanded with the eigenstates
$|\Psi_E;\Delta J\rangle$ with Hamming weight $HW(\Delta J)\ge \alpha$.
Note that if $HW(\Delta J)\ge \alpha$,
from the definition of $\pi(e)$ given in Eqn.(\ref{eq:def_pi_e})
\[(P_E)^s|\Psi_E;\Delta J\rangle=(1-1/l)^{sHW(\Delta J)}|\Psi_E;\Delta J\rangle\ .\]
From this, Inequality (\ref{lemma1_1}) immediately follows.
Inequality (\ref{lemma1_2}) can be shown by setting $s=2$ in (\ref{lemma1_1})
as the operator $P_E$ is Hermitian.\hfill$\Box$

Next we present the rigorous version of Inequality(\ref{eq:rough_ineq1}).
\newtheorem{lemma2}[theorem]{Lemma}
\begin{lemma2}
\begin{eqnarray}
\langle\Psi_V|Q|\Psi_V\rangle
&\le&1+r(r-1)\beta^{d-\alpha}F_\alpha(N)\ ,\label{eq:lemma2_1}\\
\left|\langle\Psi_V|\Psi_V\rangle-1\right|&\le& (r-1)\beta^{d-\alpha}F_\alpha(N)\ ,\label{eq:lemma2_2}
\end{eqnarray}
where,
\[
F_\alpha(N):=\sum_{i=0}^{\alpha-1}{N\choose i}(D-1)^i\ .\]
Parameter $\beta$ is defined among particle states $|e\rangle$, $|e'\rangle$
and operater $\pi(e^{\prime\prime})$, which acts on them:
\begin{equation}
\beta := \max_{\neg(e=e'=e^{\prime\prime})}
\left|\langle e|\pi(e^{\prime\prime})|e'\rangle\right|\ <1\ .
\label{eq:define_beta}
\end{equation}
The maximum value of Eqn.(\ref{eq:define_beta}) is evaluated under the
condition that the triplet $e,e',e^{\prime\prime}$ does {\it not} satisfy $e=e'=e^{\prime\prime}$.
\end{lemma2}

\noindent{\it Proof.}

Note
\begin{eqnarray}
\langle\Psi_V|Q|\Psi_V\rangle&=&\sum_i|v_i|^2\langle\Gamma_i|P_{E(A_i)}|\Gamma_i\rangle
\label{eq:lemma2proof1}\\
&&+\sum_{\{i,j,k| \neg(i=j=k)\}}
v_j^*v_k\langle\Gamma_j|P_{E(A_i)}|\Gamma_k\rangle\ .\nonumber
\end{eqnarray}
The first term on the RHS is clearly less than or equal to 1.
The second term is the sum of the cross terms generated by vectors from
different subspaces $V(E(A_i))$'s.
These should be negligible for a large enough $d$, since in that case
$V(E(A_i))$s are approximately orthogonal.

We will evaluate this second term exactly below.
From Eqn.(\ref{eq:expansion2}), we have
\begin{eqnarray}
\lefteqn{\langle\Gamma_j|P_{E(A_i)}|\Gamma_k\rangle}\label{eq:Gamma_pi_Gamma}\\
&=&\sum_{\Delta J, \Delta J'}
w^*_{j,\Delta J}w_{k,\Delta J}\ \langle\Psi_{E(A_j)};\Delta J|P_{E(A_i)}|\Psi_{E(A_k)};\Delta J'\rangle\ .
\nonumber
\end{eqnarray}
The term appearing on the RHS can be bounded from above as
\[
\left|\langle\Psi_{E(A_j)};\Delta J|P_{E(A_i)}|\Psi_{E(A_k)};\Delta J'\rangle\right|\le \beta^{d-\alpha}\ ,
\]
since the triplet $i,j,k$ does {\it not} satisfy $i=j=k$.
In addition,
because the Hamming weight of $\Delta J$ is bounded from above as
$HW(\Delta J)\le \alpha$,
there are only $F_\alpha(n)$ patterns of $w_{k,\Delta J}$
(and equivalently for $w_{j,\Delta J}$).
Thus, maximizing the RHS of Eqn.(\ref{eq:Gamma_pi_Gamma}) using the Lagrange multiplier
for $w_{j,\Delta J}$ and $w_{k,\Delta J'}$, we find
\[|\langle\Gamma_j|P_{E(A_i)}|\Gamma_k\rangle|\le F_\alpha(N)\beta^{d-\alpha}\ .\]
Applying the Lagrange multiplier again on $v_i$,
we obtain an upper bound for the second term on the RHS of (\ref{eq:lemma2proof1}),
and Inequality (\ref{eq:lemma2_1}) can be proved.
Inequality (\ref{eq:lemma2_2}) can be proved as well in a similar
manner if one notices that
$\max_{e\ne e'}|\langle e|e'\rangle|=\bar\beta\le\beta$.\hfill$\Box$

\subsubsection{Proof of Theorem 2}
From the above two lemmas
\begin{eqnarray*}
\lefteqn{\langle \Psi|Q|\Psi\rangle}\\
&\le& \langle\Psi_V|Q|\Psi_V\rangle
+2|\langle\Psi_V|Q|\Psi_\bot\rangle|+\langle\Psi_\bot|Q|\Psi_\bot\rangle\\
&\le& 1+r(r-1)\epsilon_1 + 2r\left(1+(r-1)\epsilon_1\right)\epsilon_2S +r\epsilon_2S^2
\end{eqnarray*}
and
\[\langle\Psi|\Psi\rangle=\langle\Psi_V|\Psi_V\rangle+\langle\Psi_\bot|\Psi_\bot\rangle
\ge1-(r-1)\epsilon_1+S^2\ ,\]
where
\begin{eqnarray}
\epsilon_1 &=& \beta^{d-\alpha}F_\alpha(N)\ ,\ \ 
\epsilon_2 = (1-1/l)^\alpha\ ,\label{eq:def_epsilons}\\
S&=& \langle\Psi_\bot|\Psi_\bot\rangle^{1/2}\ .\nonumber
\end{eqnarray}
({\it c.f.}, Inequalities (\ref{eq:rough_ineq1}) and (\ref{eq:rough_ineq2})),
Thus,
\begin{eqnarray*}
\lefteqn{\sum_i{\rm Pr}[A_i|\rho]\le\frac{\langle\Psi|Q|\Psi\rangle}{\langle\Psi|\Psi\rangle}}\nonumber\\
&\le&\frac{1+r^2\epsilon_1+2r\epsilon_2(1+r\epsilon_1)S+r\epsilon_2S^2}{1-r\epsilon_1+S^2}\nonumber\\
&=&r\epsilon_2+\frac{1+r^2\epsilon_1-r\epsilon_2(1-r\epsilon_1)+r\epsilon_2(1+r\epsilon_1)S}
{1-r\epsilon_1+S^2}\nonumber\\
&=:&r\epsilon_2+\frac{c_1+c_2S}{c_3+S^2}\ .
\label{theorem_proof1}
\end{eqnarray*}
Assuming $c_1,c_2,c_3>0$ and differentiating by $S$, we see that
\begin{equation}
\sum_i{\rm Pr}[A_i|\rho]\le r\epsilon_2+\frac12\frac{\sqrt{c_1^2+c_3c_2^2}+c_1}{c_3} =1+\varepsilon\ .
\label{theorem_proof2}
\end{equation}
Here, as $c_1,c_3\simeq 1$ and $c_2$ can be chosen arbitrarily small,
$\varepsilon$ on the RHS can be arbitrarily small.
\hfill$\Box$
\subsubsection{Choice of Parameters}\label{sec:choice_of_parameters}
Next,
we show that $\varepsilon$ given in (\ref{theorem_proof2}) can actually be arbitrarily small
by choosing sufficiently large $N$.
Here, we do a very rough estimate without worrying about the tightness of the bound.
First assume $r\epsilon_1\le1/2$ for $\epsilon_1$ defined in (\ref{eq:def_epsilons}), then
\[\frac{\sqrt{c_1^2+c_3c_2^2}+c_1}{c_3}\le(1+2r\epsilon_1)(2c_1+c_2)\ .\]
and immediately we have
\begin{equation}
\sum_i{\rm Pr}[A_i|\rho]\le1+4r\epsilon_2+4r^2\epsilon_1\ .
\label{eq:brob_estimate}
\end{equation}
We will prove below that the second and the third term on the RHS of Inequality (\ref{eq:brob_estimate})
can be $\le\varepsilon/2$ respectively.
As to the second term, it suffices to let $\alpha=\log_{1-1/l}\left(2^{-6}\varepsilon/r\right)$.
As to the third term, first note
\begin{equation}
F_\alpha(N)<(D-1)^\alpha\exp\left[N\cdot H_2(\alpha/N)\right]
\label{eq:ECC_formula}
\end{equation}
for $N-\alpha+1>N/D$ (see {\it e.g.}, Ref.\cite{Peterson}, Appendix A).
Using Inequality(\ref{eq:ECC_formula}), we see that
the third term on the RHS of (\ref{eq:brob_estimate}) can be bounded from above
if
\begin{equation}
N\cdot H_2\left(\frac{\alpha}{N}\right)+d\log\beta\le
\log\left(\frac{\varepsilon\beta^\alpha}{8r^2(D-1)^\alpha}\right)\ .
\label{eq:thirdterm_estimate}
\end{equation}

From a well-known theorem of the theory of error-correcting codes,
$N$ can be made arbitrarily large while keeping constant
the information rate $n'/N$ and the relative minimum distance $d/N$.
The RHS of (\ref{eq:thirdterm_estimate}) being a constant and $\log\beta$ being negative,
Inequality (\ref{eq:thirdterm_estimate}) can always be satisfied by choosing a large enough $N$.

\section{Notes for Implementation}\label{sec:implementation}
As mentioned earlier, our protocol can be implemented using currently available technology.
For example, the case of $D=2$ can be realized by single photon sources and detectors.
Probably the most familiar way of encoding is the BB84-like states that are 
often used for quantum key exchange\cite{BB84},
\[
\begin{array}{ll}
|0\rangle=\left(
\begin{array}{c}
1\\0
\end{array}
\right),&
|1\rangle=\left(
\begin{array}{c}
0\\1
\end{array}
\right),\\
|2\rangle=\frac1{\sqrt2}
\left(
\begin{array}{c}
1\\1
\end{array}
\right),&
|3\rangle=\frac1{\sqrt2}
\left(
\begin{array}{c}
-1\\1
\end{array}
\right),
\end{array}
\]
which corresponds to $q=4$, $D=2$, $l=2$, and $\beta=3/4$.
As one can see from the theory of error-correcting codes,
for sufficiently large $N$,
there exists a code $E$ with the information rate $n'/N$ that is virtually equal to 1.
Thus, in this case, a secure protocol can be constructed with $m/n\simeq 1/2$,
where $m$ is the bit number accessible by the receiver before unveiling.

For example, take $r=2^{10}$ and $\epsilon=2^{-10}$, which correspond to $\alpha\ge26$.
Then, it is guaranteed from the Gilbert-Vershamov bound (see Ref.\cite{Peterson})
that there exists a linear code with $q=4$, $N=10^5$, $d/N=10^{-2}$, and $n'/N\ge 0.95$,
for which Inequality (\ref{eq:thirdterm_estimate}) holds.
In this case, $m/n<0.53$ is satisfied.

It should be noted, however, that in real life there are problems with
information losses in optical channels and with the detection rate of photon detectors.
Consequently, Bob might end up rejecting Alice's commitment
even when an honest Alice has sent the correct commitment $|\Psi_A\rangle$.

Still, we can overcome these problems by modifying steps 3 and 4 of the open phase
(Sec.\ref{sec:proposed_protocol_procedure}) as follows.
First take an integer parameter $t>0$,
the exact value of which is determined by the noise level of the channel and the detection rate.
Then change steps 3 and 4 of the open phase as follows:
\begin{enumerate}
\setcounter{enumi}{2}
\item Bob verifies the following condition for each particle $1\le i\le N$\ :
\begin{itemize}
\item[-]
If the basis he chooses is correct, {\it i.e.}, $|e_i\rangle\in M(s_i)$,
he obtains $|e_i\rangle$ as a result of his measurement.
\end{itemize}
Then, he records as $y$ the number of particles that do {\it not} satisfy the above condition.
\item If the above condition does not hold for more than $t$ particles,
{\it i.e.}, if $y>t$, then Bob rejects the commitment.
Otherwise he accepts it.
\end{enumerate}

In other words, Bob accepts Alice's commitment even when
up to a certain number of particles does not meet the condition of step 3.
In this case too,
the security of the protocol can be shown in almost the same way as in the previous
section\cite{WorkInProgress}.

\section{Conclusion}\label{sec:conclusion}
In this paper, we have proposed a new QBSC protocol
and shown that it is secure in terms of the same security requirements as discussed in Ref.\cite{Kent}.
Our protocol has the merit that it can be implemented using currently available technology
such as single-photon sources and detectors.

An example of future work is the optimization of the protocol presented here.
As discussed in Section \ref{sec:implementation},
if the protocol is implemented using BB84-like states,
the classical error-correcting code $E$ needs to be of the order of $N\simeq 10^5$ in length.
Although possible in principle,
it is not very practical to calculate a generator matrix of this size using currently available computers.
Instead, it is much more worthwhile to optimize the protocol or the security proof
so that we can ensure the same level of security for smaller $N$.

For example, the estimate given in Section \ref{sec:choice_of_parameters} is not yet tight,
since there we were interested mainly in proving that the protocol is in fact possible.
Thus, by using a better strategy, we might be able to obtain more security for shorter bit length $n$.
Note, in particular, that even with the ratio $m/n$ of the accessible bits being fixed,
there is a variety of choices of parameters, such as $q$, $D$, and $l$.
It will be interesting to see how we can better optimize the protocol,
whether by making better choices for quantum states $|0\rangle,\dots,|q-1\rangle$
or by optimizing the classical error-correcting code $E$.

\end{document}